\documentclass{PoS}
\usepackage{graphicx,color}
\usepackage{etoolbox}

\newcommand{\MSb}{\overline{\mathrm{MS}}}
\newcommand{\MMSb}{{\rm M}\overline{\mathrm{MS}}}

\let\oldbibliography\thebibliography
\renewcommand{\thebibliography}[1]{\oldbibliography{#1}
\setlength{\baselineskip}{9.5pt}
\setlength{\itemsep}{0pt}} 

\title{Quasi-PDFs with twisted mass fermions}

\ShortTitle{Quasi-PDFs with twisted mass fermions}

\author{Constantia Alexandrou$^{ab}$,
  \speaker{Krzysztof Cichy}$^{c}$,
        Martha Constantinou$^{d}$,  
	Kyriakos Hadjiyiannakou$^{b}$, 
	Karl Jansen$^{e}$,
	Aurora Scapellato$^{c}$,
	Fernanda Steffens$^{f}$
\\
\\
        \llap{$^a$}Department of Physics, University of Cyprus, P.O. Box 20537, 1678 Nicosia, Cyprus\\
	\llap{$^b$}Computation-based Science and Technology
	Research Center, Cyprus Institute, 20 Kavafi Str.,
	Nicosia 2121, Cyprus\\
	\llap{$^c$}Faculty of Physics, Adam Mickiewicz
	University, Uniwersytetu Pozna\'nskiego 2, 61-614 Pozna\'{n}, Poland\\
	E-mail: \email{kcichy@amu.edu.pl}\\
	\llap{$^d$}Temple University, 1925 N. 12th Street,
	Philadelphia, PA 19122, USA\\
	\llap{$^e$}John von Neumann Institute for Computing (NIC), DESY, Platanenallee 6, D-15738 Zeuthen, Germany\\
	\llap{$^f$}Institut f\"{u}r Strahlen- und Kernphysik, Rheinische Friedrich-Wilhelms-Universit\"{a}t Bonn, Nussallee 14-16, 53115 Bonn
        }

\abstract{We discuss the recent progress in extracting partonic functions from the quasi-distribution approach, using twisted mass fermions. This concerns, among others, the investigation of several sources of systematic effects. Their careful analysis is a prerequisite to obtain precise determinations of PDFs from the lattice with realistic estimates of all uncertainties.
In these proceedings, we shortly discuss systematic effects in the matching procedure.
Moreover, we present preliminary results from our new simulations at the physical point. They involve, additionally, the dynamical strange and charm quarks, as well as a larger volume and a smaller lattice spacing than in our previous computations. In addition, we show first results from computations of generalized parton distributions (GPDs) in the quasi-distribution framework.}

\FullConference{37th International Symposium on Lattice Field Theory - Lattice2019\\
		16-22 June 2019\\
		Wuhan, China}

\begin{document}

\section{Introduction}
\vspace*{-4mm}
\noindent Recent years have witnessed a tremendous progress in calculations of partonic distribution functions from first principles.
For a long time, lattice computations were restricted to the lowest 2-3 moments of parton distribution functions (PDFs) or generalized parton distributions (GPDs).
The situation changed with the advent of Ji's proposal \cite{Ji:2013dva} to calculate quasi-distributions, objects related to the desired light-cone distributions and computable on the lattice.
Quasi-distributions share the same infrared physics with their light-cone counterparts and the difference in the ultraviolet can be subtracted using perturbation theory, in the so-called matching procedure.
The approach has been very vigorously explored theoretically and numerically, see e.g.\ Refs.~\cite{Xiong:2013bka,Lin:2014zya,Alexandrou:2015rja,Chen:2016utp,Chen:2016fxx,Alexandrou:2016jqi,Briceno:2017cpo,Ishikawa:2017faj,Ji:2017oey,Constantinou:2017sej,Alexandrou:2017huk,Ji:2017rah,Wang:2017qyg,Green:2017xeu,Stewart:2017tvs,Izubuchi:2018srq,Alexandrou:2018pbm,Briceno:2018lfj,Spanoudes:2018zya,Radyushkin:2018nbf,Karpie:2018zaz,Liu:2018uuj,Lin:2018qky,Alexandrou:2018eet,Braun:2018brg,Karpie:2019eiq,Alexandrou:2019lfo,Cichy:2019ebf} for published studies concerning PDFs of the nucleon, including renormalizability, concrete renormalization prescriptions, matching, nucleon mass corrections, finite volume effects, higher-twist effects and lattice simulations with non-physical and finally physical quark masses.
The quasi-distribution approach was also applied to other hadrons, see e.g.\ Refs.~\cite{Chen:2018fwa,Izubuchi:2019lyk,Chai:2019rer}, and other direct approaches for extracting the $x$-dependence of partonic functions were also pursued, e.g.\ Refs.~\cite{Radyushkin:2017cyf,Orginos:2017kos,Ma:2017pxb,Bali:2018spj,Sufian:2019bol,Joo:2019jct,Radyushkin:2019owq,Joo:2019bzr}. 
For an extensive review of recent activities in this field, we refer to Refs.\ \cite{Cichy:2018mum,Monahan:2018euv}.

\vspace*{-5mm}
\section{Theoretical and lattice setup}
\vspace*{-4mm}
\noindent Starting with the renormalized matrix elements $h_\Gamma^{\MMSb}(z,P_3)= Z^{\MMSb}(z)\langle P\vert \, \overline{\psi}(z)\,\Gamma W(z,0)\,\psi(0)\,\vert P\rangle$ (where the nucleon momentum is $P=(P_0,0,0,P_3)$, $Z^{\MMSb}(z)$ is the renormalization function in the $\MMSb$ scheme \cite{Alexandrou:2019lfo}, $W(z,0)$ is the Wilson line of length $z$ and $\Gamma$ determines the type of PDF), one obtains quasi-PDFs as the Fourier transform with respect to $z$, $\tilde{q}^{\MMSb}(x,P_3)=\hspace*{-0.1cm}\int_{-\infty}^{+\infty}\hspace*{-0.1cm}\frac{dz}{4\pi}\,e^{-ixP_3z}\,h_\Gamma^{\MMSb}(z,P_3)$.
We perform the computations for one ensemble of configurations, with $N_f=2$ maximally twisted mass quarks with a clover term (MTM+clover), at the physical pion mass, with lattice size of $48^3\times 96$ and lattice spacing $a=0.0938(3)(2)$~fm.
The nucleon is boosted to $\frac{6\pi}{L}$, $\frac{8\pi}{L}$ and $\frac{10\pi}{L}$ ($0.83,\,1.11,\,1.38$~GeV).
The three-point functions entering the matrix elements are computed at four source-sink separations ($t_s=0.75$, $0.84$, $0.93$ and $1.12$ fm).
The number of measurements at the largest $t_s$ is 9600, 38250 and 72990 measurements (for our three boosts, respectively).
For more details about the computation of the bare matrix elements and their renormalization functions, we refer to Ref.~\cite{Alexandrou:2019lfo}. 

Thus computed quasi-PDFs differ from light-cone PDFs only in the UV region and this difference can be accounted for in perturbation theory, using a matching procedure.
In this work, we use the matching formulae developed in Refs.~\cite{Alexandrou:2018pbm,Alexandrou:2019lfo}, for which the input quasi-PDF is expressed in the $\MMSb$ scheme.
The final step is to apply nucleon mass corrections, derived in Ref.~\cite{Chen:2016utp}.

\vspace*{-5mm}
\section{Systematic effects in matching}
\vspace*{-4mm}
In Ref.~\cite{Alexandrou:2019lfo}, we analyzed several systematic effects that appear at different stages of the computation.
Here, we briefly discuss the ones at the level of the matching procedure, for the unpolarized PDFs case.
However, the magnitudes of the considered effects in helicity and transversity PDFs were found to be very similar.

\begin{figure}[t!]
\begin{center}
\includegraphics[scale=0.68]{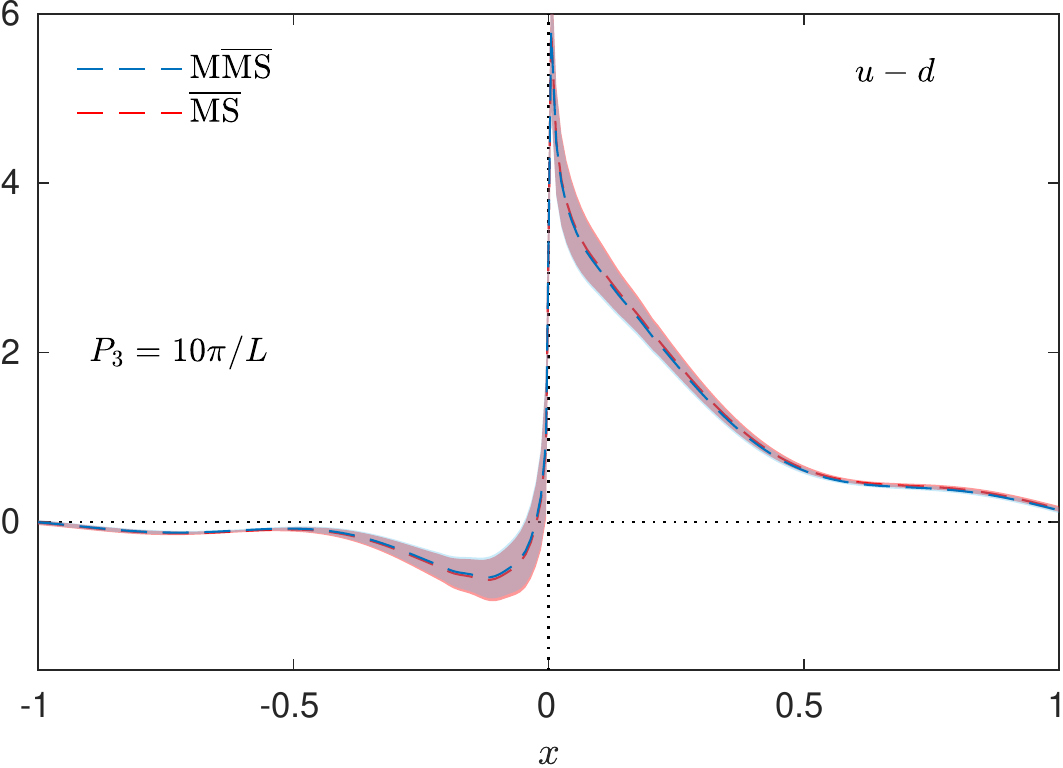}\,\,\,
\includegraphics[scale=0.68]{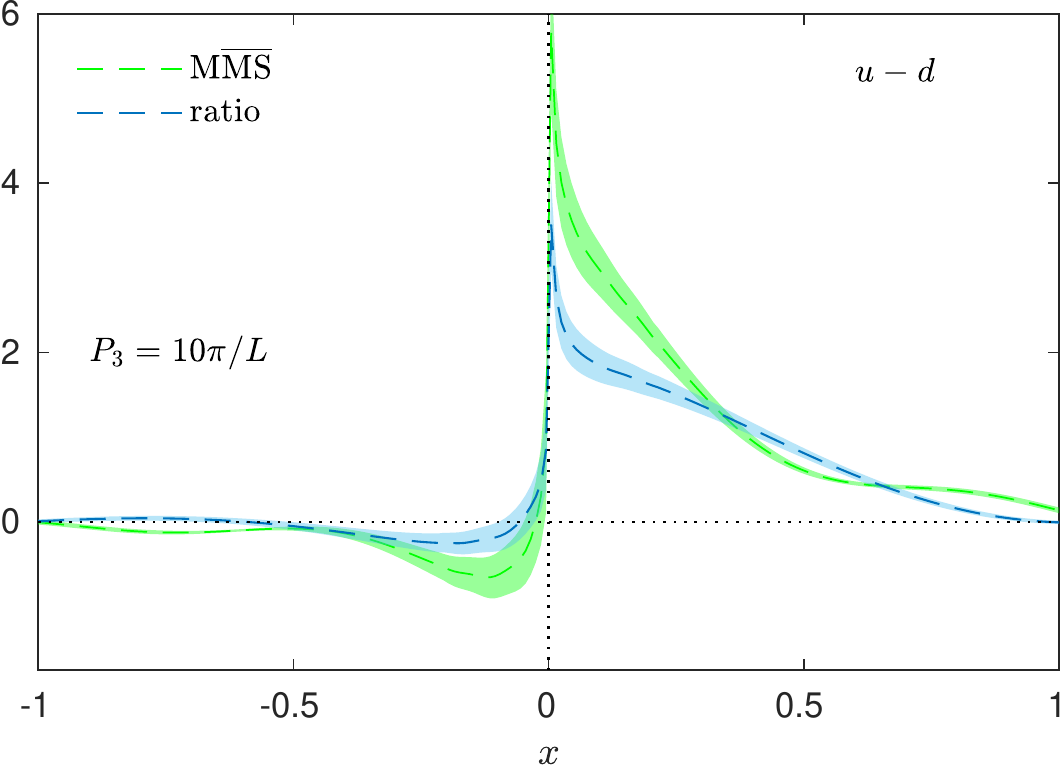}\\
\hspace*{0mm}\includegraphics[scale=0.66]{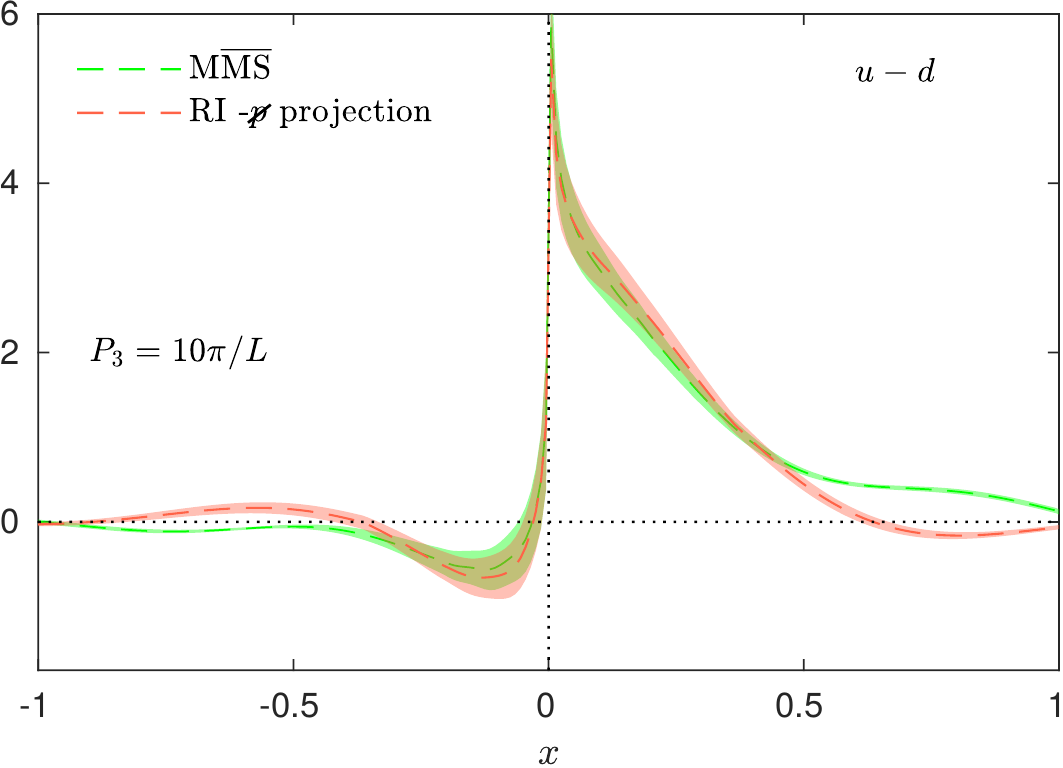}
\includegraphics[scale=0.77]{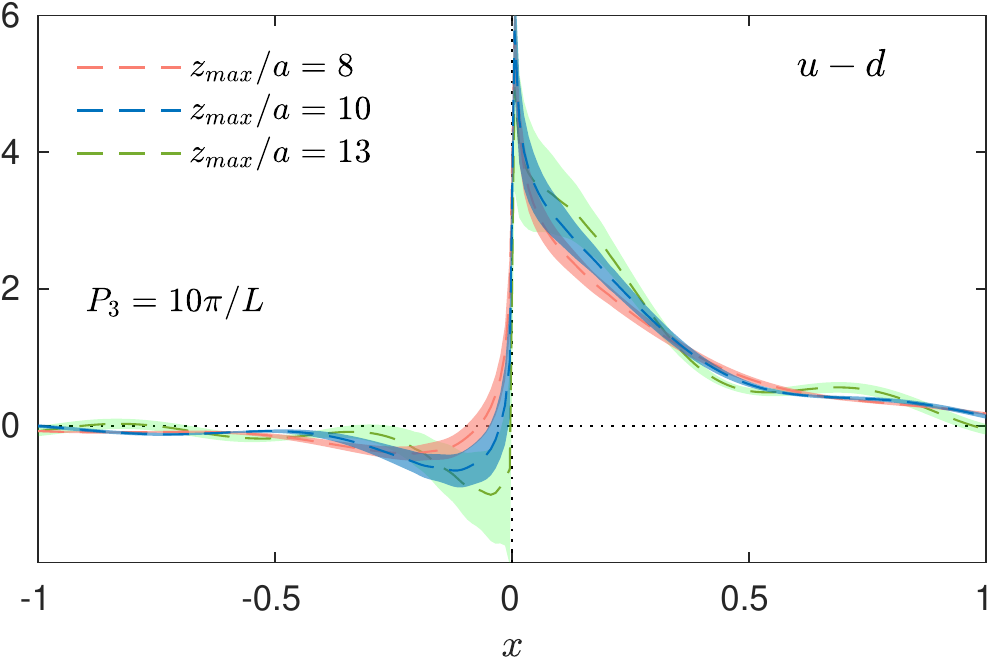}
\vspace*{-2mm}
\caption{Selected tests of the matching procedure for the unpolarized PDF ($P_3=10\pi/L$). We show a comparison of the matched PDF in the $\MMSb$ scheme and the $\MSb$ scheme (top left), the ratio scheme (top right) and the RI scheme (bottom left). The bottom right plot illustrates the dependence of the $\MMSb\rightarrow\MSb$ matched PDF on the choice of the maximum $z/a$ used in the Fourier transform.}
\label{fig:matching}
\end{center}
\end{figure}

In Fig.~\ref{fig:matching} (upper left), we show the effect of taking into account the conversion factor between the $\MSb$ scheme and the $\MMSb$ scheme.
In Refs.~\cite{Alexandrou:2018pbm,Alexandrou:2018eet}, we used the matching formulae that require the input quasi-PDF renormalized in the $\MMSb$ scheme, but the conversion factor $\MSb\rightarrow\MMSb$ was then unavailable.
An explicit test of the effect of this conversion factor (derived in Ref.~\cite{Alexandrou:2019lfo}) confirmed the expectation that it is small, roughly an order of magnitude smaller than statistical errors.
In the upper right and lower left panels of Fig.~\ref{fig:matching}, we compare the matched PDFs obtained from $\MMSb\rightarrow\MSb$ matching formulae \cite{Alexandrou:2019lfo} and from ratio$\rightarrow\MSb$ \cite{Izubuchi:2018srq} and RI$\rightarrow\MSb$ matching procedures \cite{Stewart:2017tvs}.
All these formulae are equivalent, up to neglected higher-order perturbative effects.
Since all of them are calculated only to one-loop level, the observed differences can be attributed to truncation effects.
Hence, we find these are rather significant, both at low-$x$ ($\MMSb$ vs.\ ratio) and at high-$x$ ($\MMSb$ vs.\ RI).
This necessitates the calculation of at least two-loop matching.
Finally, we compare the effects of truncating the Fourier transform at different maximal lengths of the Wilson line, $z_{\rm max}/a$.
We observe that all reasonable choices for $z_{\rm max}$ lead to compatible results.
However, as discussed in Ref.~\cite{Karpie:2019eiq}, the truncated and discretized Fourier transform is ill-defined and advanced reconstruction methods need to be applied in the future.

\begin{figure}[t!]
\begin{center}
\includegraphics[scale=0.47]{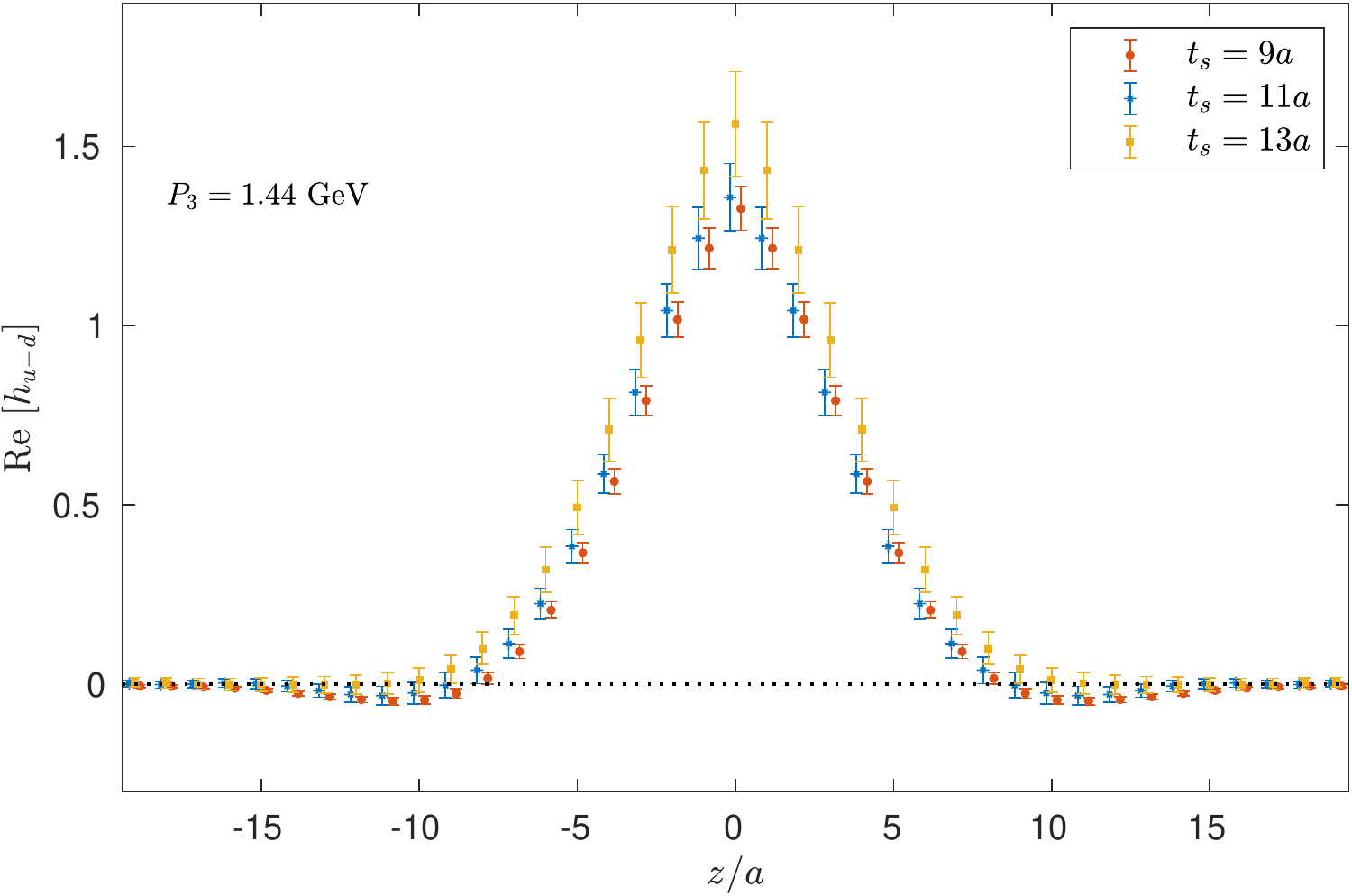}\,\,\,
\includegraphics[scale=0.47]{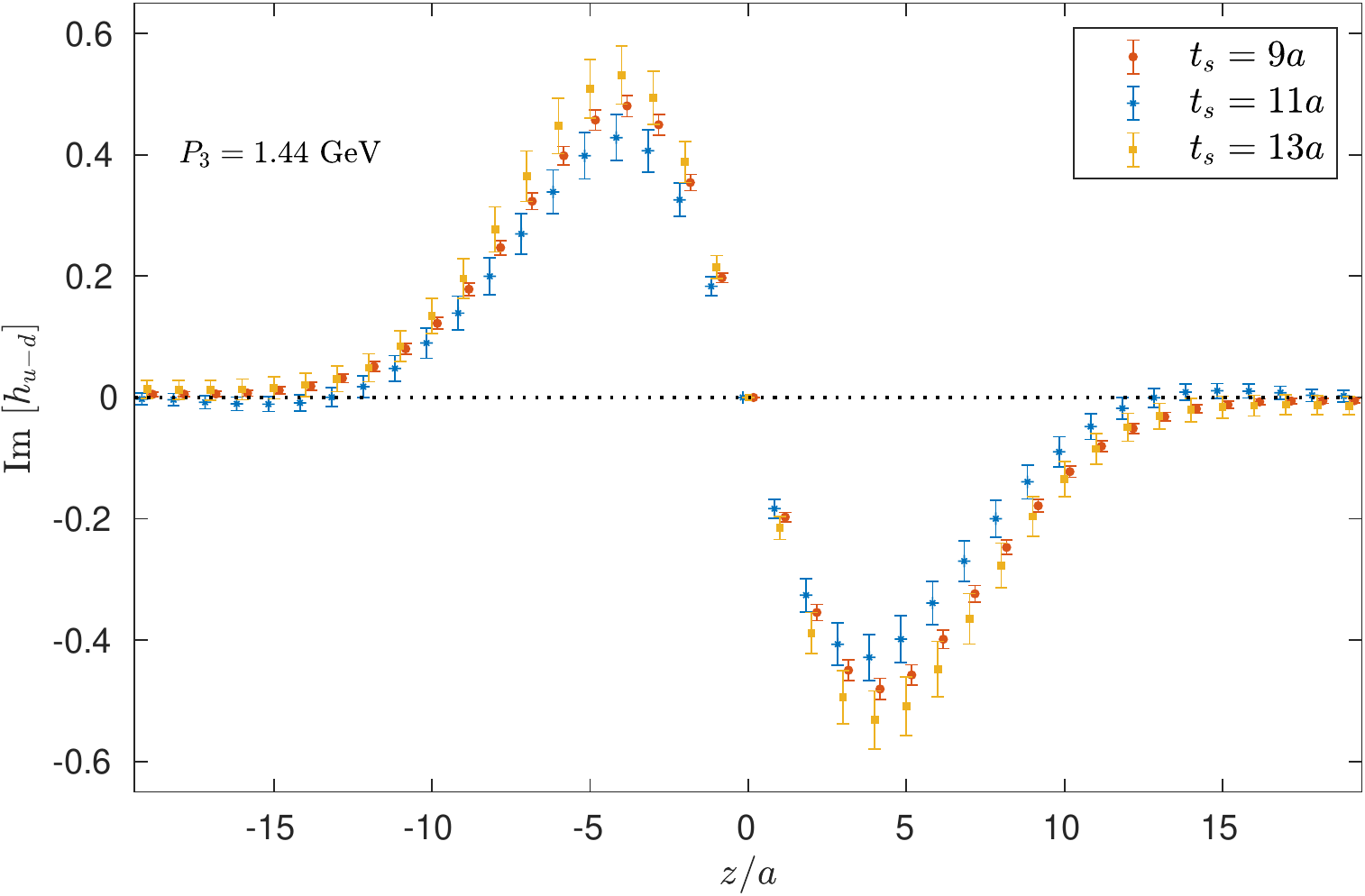}
\vspace*{-2mm}
\caption{Preliminary results for the real (left) and imaginary (right) part of matrix elements for unpolarized quasi-PDFs. Three source-sink separations are used. Nucleon boost is $1.44$~GeV. Ensemble: $N_f=2+1+1$ MTM+clover quarks, physical pion mass, lattice size $64^3\times128$, $a=0.081$~fm.}
\label{fig:64c}
\end{center}
\end{figure}

\vspace*{-5mm}
\section{Preliminary results for an $N_f=2+1+1$ ensemble}
\vspace*{-4mm}
One of our recent research directions involves computations of quasi-PDFs for another ensemble of configurations, with $N_f=2+1+1$ MTM+clover quarks at the physical pion mass, larger volume  $64^3\times 128$ and a smaller lattice spacing, $a=0.081$~fm. 
In Fig.~\ref{fig:64c}, we present results at the nucleon boost $P_3=12\pi/L\approx1.44$~GeV, obtained from 31770 measurements, and at three values of $t_s=0.73, 0.89, 1.05$~fm.
We do not observe significant excited states effects -- for most values of $z/a$, in both the real and imaginary parts, matrix elements are compatible for all source-sink separations.
Exceptions are small tensions at low-$z$ between $t_s=9a$ and $t_s=13a$ (real part) and at intermediate-$z$ for $t_s=11a,13a$ (imaginary part).
We are currently increasing statistics for the largest two separations to clarify these effects and obtain statistical precision at the level of the one for $t_s=9a$.

\begin{figure}[t!]
\begin{center}
\includegraphics[scale=0.88]{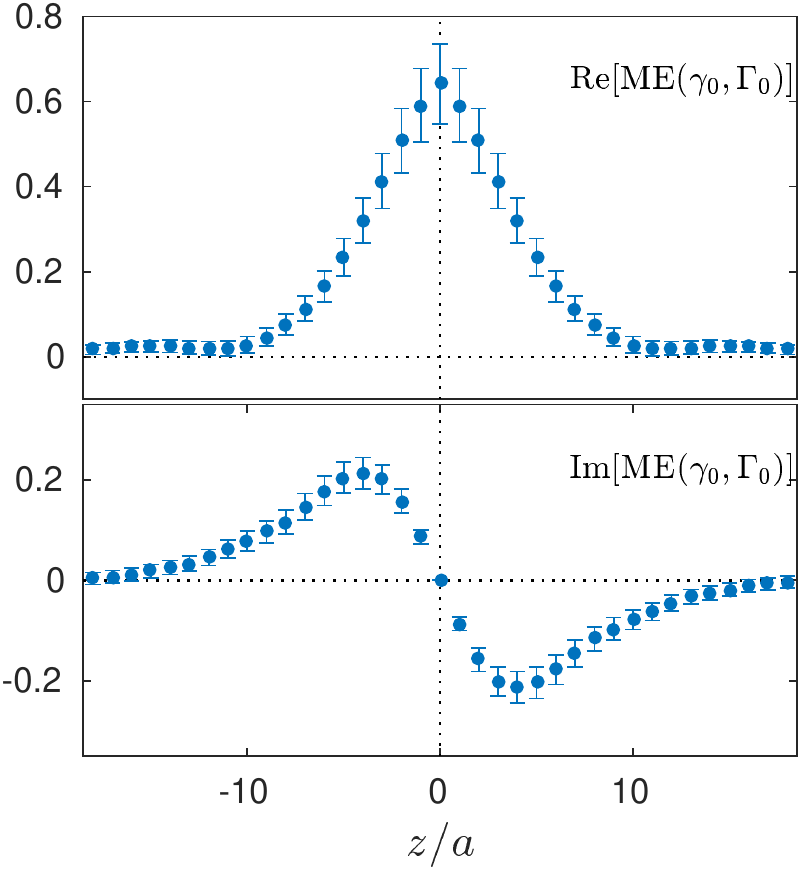}\,\,\,
\includegraphics[scale=0.88]{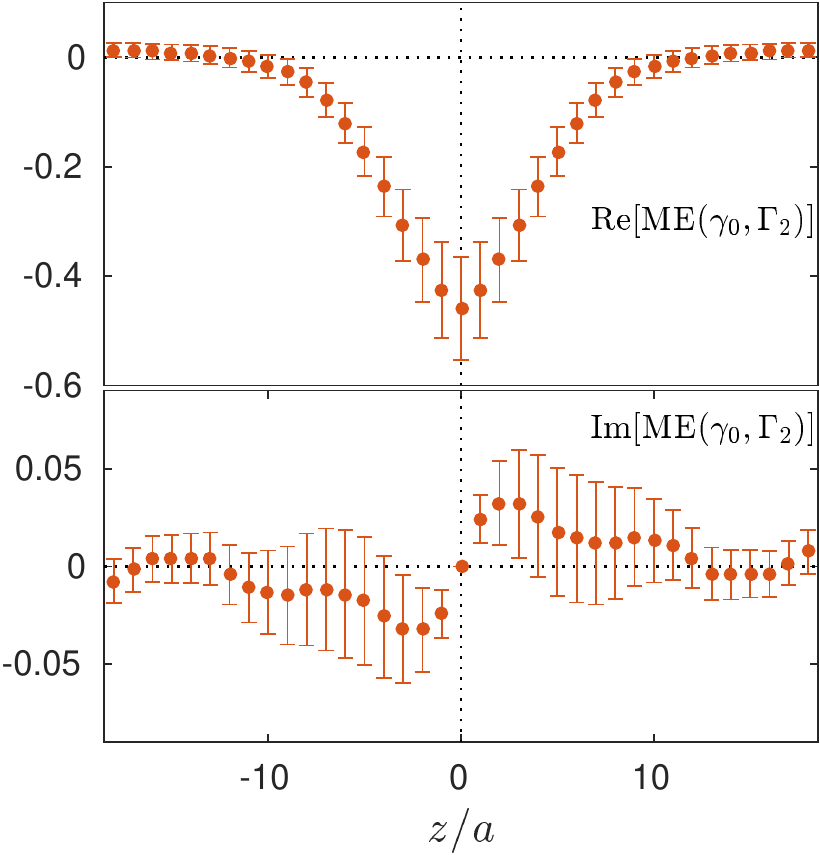}
\vspace*{-2mm}
\caption{Preliminary results for the matrix elements for unpolarized quasi-GPDs. The upper/lower plots show the real/imaginary part. The left/right plots correspond to unpolarized/polarized projectors defined in the text. The average nucleon boost is $0.83$~GeV, $\xi=0$, $Q^2=0.69$~GeV$^2$. Ensemble: $N_f=2+1+1$ MTM+clover quarks, $m_\pi\approx270$~MeV, lattice size $32^3\times64$, $a=0.094$~fm.}
\label{fig:GPDs-ME}
\end{center}
\end{figure}

\vspace*{-5mm}
\section{Preliminary results for quasi-GPDs}
\vspace*{-4mm}
Another direction that we have been recently exploring is the computation of GPDs using the quasi-distribution approach\footnote{See also Ref.~\cite{Chen:2019lcm} for exploratory work about quasi-GPDs in the pion.}.
Quasi-GPDs are defined analogously to quasi-PDFs, but they involve momentum transfer, $t=-Q^2$, between the initial and final hadron states.
The Fourier transform of the underlying matrix elements is a combination of two (four) quasi-GPDs in the chiral-even (odd) case.
In the chiral-even case, these are commonly denoted as $H$ and $E$ quasi-GPDs.
They are functions of the Bjorken-$x$, the renormalization scale, the momentum transfer and the so-called quasi-skewness variable, $\xi=-Q_3/2P_3$, where a boost along the $3$-direction is assumed and $P_3$ is the average momentum.
Using different projectors, one can disentangle the two (four) contributions and subject them to matching to obtain their light-cone counterparts.
For zero quasi-skewness, the matching is identical to the one for quasi-PDFs, while $\xi\neq0$ implies different matching, as derived in Ref.~\cite{Liu:2019urm}.
Physically, $\xi\neq0$ implies the existence of the so-called ERBL region ($|x|<\xi$).

\begin{figure}[t!]
\begin{center}
\includegraphics[scale=0.635]{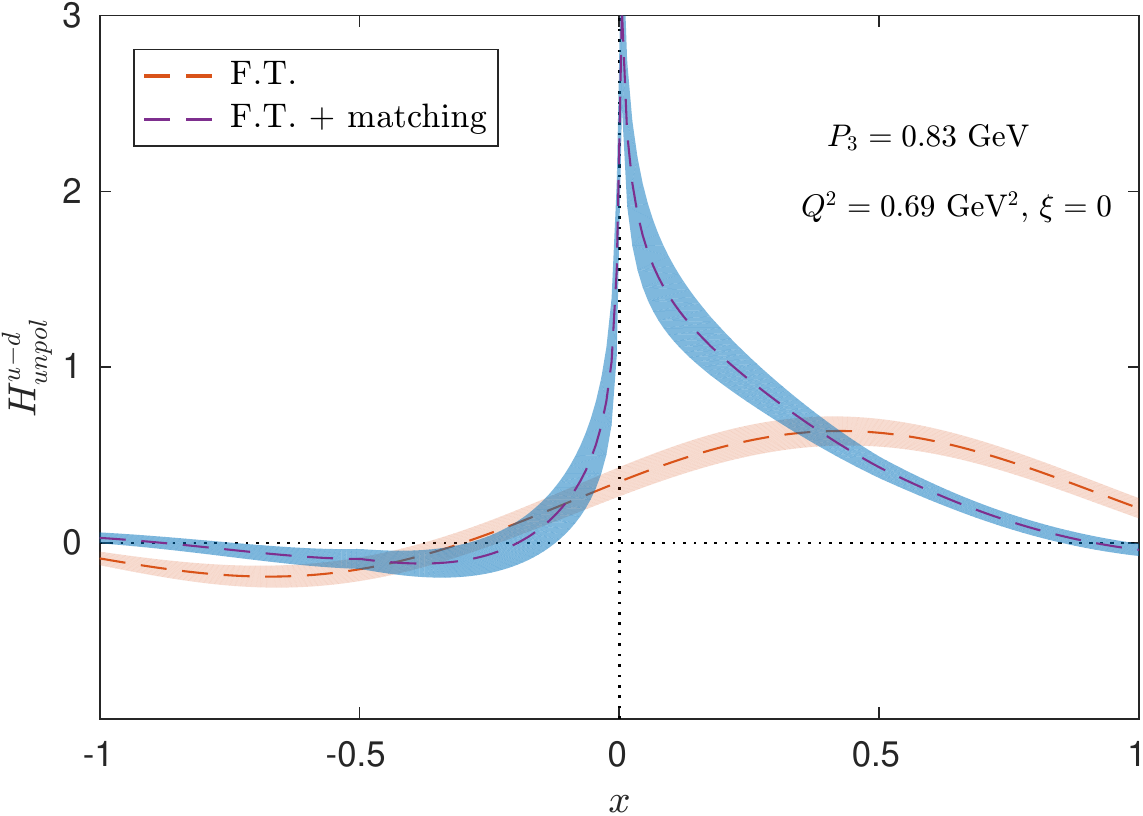}\,\,\,
\includegraphics[scale=0.676]{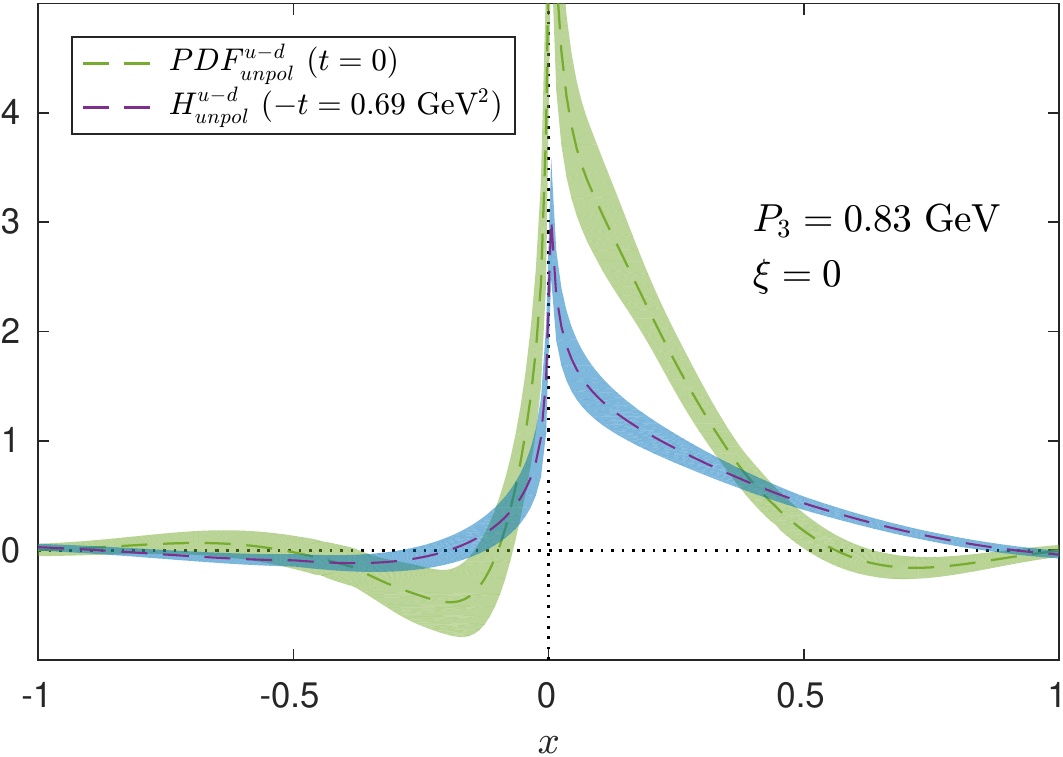}
\vspace*{-2mm}
\caption{Left: the unpolarized $H$ quasi-GPD and the matched $H$-GPD. Right: comparison of the unpolarized matched PDF and the $H$-GPD. The average nucleon boost is $0.83$~GeV, for GPD: $\xi=0$, $Q^2=0.69$~GeV$^2$. Ensemble: $N_f=2+1+1$ MTM+clover quarks, $m_\pi\approx270$~MeV, lattice size $32^3\times64$, $a=0.094$~fm.}
\label{fig:GPDs}
\end{center}
\end{figure}

Our lattice setup for this exploratory study involves $N_f=2+1+1$ MTM+clover quarks at $m_\pi\approx270$~MeV, lattice size  $32^3\times 64$ and $a=0.094$~fm.
We compute matrix elements for several nucleon boosts and momentum transfers, using $t_s\leq12a$.
In Fig.~\ref{fig:GPDs-ME}, we show bare matrix elements for unpolarized GPDs (920 measurements) at $P_3=0.83$~GeV ($\xi=0$, $Q^2=0.69$~GeV$^2$), using the unpolarized ($\Gamma_0\equiv(1+\gamma_0)/4$) and polarized ($\Gamma_2\equiv(1+\gamma_0)i\gamma_5\gamma_2/4$) projectors that allow us to disentangle the $H$ and $E$ functions.
The left plot of Fig.~\ref{fig:GPDs} shows the Fourier transform of the RI-scheme renormalized matrix element corresponding to the $H$ quasi-function and the matched $H$-GPD.
In the right plot, a comparison of this GPD with the corresponding matched PDF obtained from the same ensemble and with the same nucleon boost.
It is clear that the GPD is suppressed with respect to the PDF, as expected.
The $x$-integral of the GPD gives the Pauli form factor $F_1^{u-d}$ and the integral value we obtain, $0.62(8)$, is compatible with our renormalized matrix element (of the $H$ function) at $z=0$ ($0.61(8)$) and independent direct lattice extractions using local matrix elements at similar values of $Q^2$.
We are currently increasing statistics and computing various kinematic setups to obtain a large coverage of the $(P_3,t,\xi)$ parameter space.

\vspace*{-4mm}
\section*{Acknowledgments}
\vspace*{-4mm}
\setlength{\baselineskip}{4mm}
\begin{footnotesize}
\noindent K.C.\ is supported by the National Science Centre grant SONATA BIS no.\ 2016/22/E/ST2/00013.
This work has received funding from the European Union's Horizon 2020 research and innovation
programme under the Marie Sk\l{}odowska-Curie grant agreement no.\ 642069 (HPC-LEAP). F.S.\
is funded by the Deutsche Forschungsgemeinschaft (DFG) project no.\ 392578569. 
M.C.\ acknowledges financial support by the U.S. Department of Energy (DoE), Office of Nuclear Physics, within the framework of the TMD Topical Collaboration, by the National Science Foundation under Grant No.\ PHY-1714407 for the work on quasi-PDFs, and by DoE under Grant No. DE-SC0020405 (Early Career) for the quasi-GPDs.
This research used resources of the Oak Ridge Leadership Computing Facility, which is a DOE Office of Science User Facility supported under contract DE-AC05-00OR22725, J\"ulich Supercomputing Centre, Prometheus supercomputer at the Academic Computing Centre Cyfronet AGH in Cracow, Okeanos supercomputer at the Interdisciplinary Centre for Mathematical and Computational Modelling in Warsaw, Eagle supercomputer at the Poznan Supercomputing and Networking Center.
Computations for the work on quasi-GPDs were carried out in part on facilities of the USQCD
 Collaboration, which are funded by the Office of Science of the U.S.\
Department of Energy.
\end{footnotesize}

\vspace*{-5mm}
\bibliographystyle{JHEP}
\bibliography{references}

\end{document}